\documentclass[11pt]{article}
\usepackage{latexsym}
\usepackage{amssymb}
\usepackage{graphics}
\usepackage{epsfig}
\usepackage{rotating}
\usepackage{amsfonts}
\usepackage{amsmath}
\usepackage{hyperref}
\hypersetup{
	colorlinks=true,
	linkcolor=blue,
	filecolor=magenta,
	urlcolor=cyan,
}
\DeclareMathOperator{\asinh}{asinh}
\begin{document}

\begin{titlepage}
\begin{flushright}
IRMP-CP3-24-25\\
\end{flushright}

\vspace{5pt}

\begin{center}

{\Large\bf Newtonian Gravity and Galaxy Rotation Curves:}\\

\vspace{7pt}

{\Large\bf An Axisymmetric Green's Function Perspective}\\

\vspace{40pt}

Jan Govaerts$^{a,b,c}$

\vspace{30pt}

$^{a}${\sl Centre for Cosmology, Particle Physics and Phenomenology (CP3),\\
Institut de Recherche en Math\'ematique et Physique (IRMP),\\
Universit\'e catholique de Louvain (UCLouvain),\\
2, Chemin du Cyclotron, B-1348 Louvain-la-Neuve, Belgium}\\
E-mail: {\em Jan.Govaerts@uclouvain.be}\\
ORCID: {\tt http://orcid.org/0000-0002-8430-5180}\\
\vspace{15pt}
$^{b}${\sl International Chair in Mathematical Physics and Applications (ICMPA--UNESCO Chair)\\
University of Abomey-Calavi, 072 B.P. 50, Cotonou, Republic of Benin}\\
\vspace{15pt}
$^{c}${\sl Fellow of the Stellenbosch Institute for Advanced Study (STIAS),\\
Stellenbosch, Republic of South Africa}\\

\vspace{10pt}


\vspace{20pt}

\begin{abstract}
\noindent
The standard proposal within the context of General relativity and its weak field Newtonian limit for the nature of dark matter is that it
consists of dark matter particles of unknown type. In the present work and specifically for spiral galaxy rotation curves, an alternative possibility is explored, in the form
of an axially symmetric vortex mass distribution of finite extent threading the centre of the galaxy and perpendicular to its disk.
Some general considerations are developed and characteristic properties are identified, pointing to the potential interest of such an alternative to be studied
in earnest.

\end{abstract}

\vspace{60pt}

\hspace{30pt} Dark matter; Axisymmetric distribution

\end{center}

\end{titlepage}

\setcounter{footnote}{0}

\section{Introduction and Motivation}
\label{Intro}

Consider an electrically charged point particle, or even a whole sample of such point charges whose mutual inter-actions are ignored, to be used all as
so many independent probes for electromagnetic field configurations in vacuum, beginning say with some static electric field devoid of a magnetic component.
And imagine that all these point charges are observed to be following circular trajectories all sharing a common centre and
 all confined essentially to a single planar disk, while displaying furthermore the most distinctive feature of a flat rotation curve. Namely, to be
 such that the speed $v(\rho)$ of each trajectory is independent of its distance $\rho$ (and thus its radius) measured to its centre within the disk,
 namely $v(\rho)=v_0$, at least for distances larger than some minimal one, $\rho\ge \rho_{\rm min}$.
 
 Steadfast in the conviction that Maxwell's laws as such apply to all classical electromagnetic phenomena, and provided the speed $v_0$ be sufficiently nonrelativistic
 to warrant recourse to Newton's dynamics, the kinematics of circular motion implies of course that the electric field responsible for such centripetal
 acceleration $v^2(\rho)/\rho=v^2_0/\rho$ that varies as $1/\rho$, possesses itself such a dependency in the distance $\rho$ to the centre of the disk.
 
 As is well known, on account of Gauss' law
 such an electric field configuration may result from an infinitely long uniformly charged straight line of linear charge density
 $\lambda$, perpendicular to the disk and threading its centre, with its transverse radial component then taking the value
 \begin{equation}
 E_\rho(\rho)=\frac{\lambda}{2\pi\epsilon_0}\,\frac{1}{\rho}.
 \label{eq:Erho}
 \end{equation}
 With $q$ and $m$ being the charge and the mass of a point charge, respectively,
 under the condition that $\lambda q<0$ such a field configuration indeed allows for circular trajectories in any plane perpendicular to the charged line, of which
 the speed is given as $v^2(\rho)=|\lambda q|/(2\pi\epsilon_0\,m)=v^2_0$ independently of the value for $\rho$.
 
 Thus even though such a configuration for the electric field does not explain why circular trajectories would have to be confined to a specific single disk perpendicular
 to the charged straight line (indeed such a field remains invariant under translations along that line), yet the geometry of
 the probe charges' trajectories provides strong support to the argument that the observation of a flat rotation curve for their circular trajectories manifests
 the presence of a straight line of charge threading the centre of their confining rotating disk.

Incidentally let us remark that the electrostatic scalar potential related to the electric field~(\ref{eq:Erho}) and sharing its cylindrical and translational symmetries
is represented by
 \begin{equation}
 \Phi_\lambda(\rho)=\frac{\lambda}{2\pi\epsilon_0}\ln(\rho/\rho_0),
 \label{eq:Phirho}
 \end{equation}
 with $\rho_0$ being an arbitrarily chosen finite reference value for the distance as measured to the straight charged line, at which that potential is taken to vanish.
 Besides the necessary behaviour proportional to $\ln\rho/\rho_0$ of the scalar potential produced by such a charge distribution leading to a flat rotation curve,
 note that it can never asymptotically vanish at infinity in all possible spacelike directions -- in fact quite to the contrary, since it asymptotically grows infinite
 in all spacelike directions, except when $\rho=\rho_0$, a feature which certainly remains unavoidable since the (idealised) source straight line charge distribution
 itself extends up to an infinite distance in two opposite directions.

For sure, but now rather purely within a gravitational context, flat velocity rotation curves of rotating disks
are a genuinely observed landmark and distinctive feature of spiral galaxies certainly, and to some lesser extent
possibly even of elliptic galaxies (for some reviews and references to the original literature, see Refs.\cite{DarkWiki,Dark1,Dark2,Faber,Rubin,Strigari}).
For a large sample of spiral galaxies as available nowadays \cite{Sofue1, Sofue2, Sofue3, Sofue4},
and starting at distances from their centre still less than the radius
of their electromagnetically visible outer edge, essentially flat rotation curves are observed to reach outwards over distances manifold times larger than
that outer edge radius -- namely at distances of at least a few tens of kpc's and more --, with speed values $v_0$ ranging mostly between some 150~km/s
to some 300~km/s (see for instance Figure~1 in Refs.\cite{Sofue1,Sofue3}), say some 230 km/s on average, namely a value which remains well within
the nonrelativistic realm with $v_0/c < 10^{-3}$.

At the smaller distances where the rotation curve is not essentially flat yet,
the details of the behaviour of the speed $v(\rho)$ as function of distance are dependent on the finer structures of the distribution of electromagnetically visible
thus known matter within the galaxy, namely its bulge (which quite generally encloses a central supermassive black hole) and the properties of its
(possibly double thick and thin) disk components (while a close to spherical stellar and gaseous halo beyond the disk radius contributes as well but generally
to a significant lesser degree)~--~and this of course by ignoring the finer coarse grained structures related to the galaxy's spiral arms and their proper dynamics.
However the close to universal character of the value $v_0$ of the speed of flat rotation curves, as well as the large distances required before possibly witnessing
finally a -- not yet observed -- turn-around and decrease in rotation curves, both call for a more encompassing understanding independent of the details
of the electromagnetically visible matter distribution proper to each galaxy, thereby possibly holding a clue to a physical rationale
for the existence of such flat rotation curves in the gravitational context of galaxy dynamics and their wider clustering at even larger scales still. 

Given the nonrelativistic speed value $v_0\simeq 230$~km/s as well as the extra-galactic distance scales of at least
a few tens of kpc's involved in these flat galaxy rotation curves,
clearly the corresponding small centripetal accelerations of some $10^{-11}-10^{-10}$~m/s$^2$ manifest the existence at those distances
of a weak gravitational field typically possessing such a value. Under such circumstances and as is totally warranted for a first-order assessment within this context,
Newton's universal law for gravity provides a perfectly well justified approximate representation of the manifestation of the actual laws and gravitational consequences
of Einstein's Theory of General Relativity for the curvature of space-time in the presence of energy-matter distributions, at least for static massive sources.

Steadfast in the conviction that Einstein's laws of General Relativity apply as such to all classical gravitational phenomena, and given the nonrelativistic dynamics
involved, once again a recourse to Newton's dynamics inclusive of his law for gravity -- mathematically parallel to Coulomb's law -- is warranted,
which allows to transcribe {\sl mutatis mutandis} to the gravitational realm now the considerations outlined above in the case of flat rotation curves
in the electromagnetic realm. Thereby thus raising the question: which type of mass distribution is manifested by the observed flat galaxy rotation curves?

Historically ever since the first such measurements of galaxy rotation curves were attempted, it quickly became apparent that these curves
reveal the necessary existence of some form of yet unobserved matter, or invisible ``missing" mass, partaking in the making up of galaxies and contributing
to their dynamics, and beyond these as well. Thus a form of ``dark matter" invisible through the electromagnetic interaction and yet necessarily present on all astrophysical
and cosmological scales and dynamically so crucial indeed to structure formation through the gravitational interaction acting on all these scales:
the dark matter mystery of the Universe, with the first observational hints of this gravitational conundrum dating back to almost a century ago,
and its systematic exploration and study being pursued and developed vigourously over the last five decades at least.

One suggestion natural enough on first thought and inspired by Gauss' law, is to imagine that dark matter is provided by distributions of collisionless
freely streaming gravitationally only interacting dark matter particles of yet unknown nature. Flat galaxy rotation curves may then be accounted for through
a close to spherically symmetric halo of dark matter particles whose radius is much larger than that of the electromagnetically visible embedded galaxy,
and whose radial distribution is tailored to reproduce the observed flat rotation curve for any given galaxy. By extension to the larger astrophysical and
cosmological scales, the same strategy is considered in terms of dark matter particle distributions whose configurations are adapted to the structures
and dynamics to be accounted for. Under the assumption that Newton's and Einstein's laws for the gravitational interaction do not require any extension
neither at the largest scales nor at the smallest accelerations, this suggestion of the existence of dark matter particles remains by far the most prevailing
hypothesis towards an eventual physical understanding of the dark matter mystery \cite{DarkWiki,Dark1,Dark2,Faber,Rubin,Strigari}.

Therefore the last decades have witnessed many original instrumental and technical developments, experimental strategies, and theoretical model constructions
involving manifold known and unknown candidate elementary particles (including the possibility of primordial black holes), indeed of ever increasing true inventiveness
and ingenuity, going hand in hand and aiming towards the direct or indirect detection and identification finally of some possible dark matter particle candidate.
But to no avail up to now.

On the other hand, and this still within the specific context of Newtonian and Einsteinian gravity, by borrowing from the electromagnetic analogy outlined above
an alternative approach suggests itself obviously
quite naturally as well. Namely that flat galaxy rotation curves manifest the presence of some invisible mass distribution yet still
of some unknown nature, however not in the form of a halo of freely streaming dark matter particles but rather in the form of some dark matter distribution structured
along a line threading the centre of the galaxy's disk, perpendicular to it, of finite transverse width, and of a finite linear extent much greater -- possibly on the order
of average  inter-galactic distances -- than the radius of that disk. For lack of a better choice (and to avoid the use of the term ``string" which suggests, among other
meanings, a structure of vanishing width) let us refer herein to such a structure of finite length as a ``(cosmic) dark matter vortex'', or ``vortex'' for short
(the possibility that this structure may be rotating on itself is left open in the present discussion).

The suggestion that infinitely long straight cosmic strings of vanishing transverse width and of finite uniform linear mass distribution -- as a fair first-order
idealised approximation relevant at galactic distance scales for a more realistic configuration -- could explain
the mystery of flat galaxy rotation curves is not new, but was rapidly dismissed for the following reason. Invariance of such a gravitational
configuration under translations along the cosmic string axis implies in effect a dimensional reduction of 3+1 dimensional General Relativity to 2+1 dimensions,
with a simple point mass as the source of the gravitational field in 2+1 dimensions. As is well known \cite{Deser}, given the restriction of asymptotic flatness of the
2+1 dimensional metric, in spite of the presence of the massive point source the corresponding space-time then remains flat with thus no gravitational pull whatsoever
acting on any point test mass. As a matter of fact the sole manifestation of the presence of the massive point source is the existence of a conical angular deficit for
closed geodesics surrounding the point source (in other words the dependency of the space-time metric on the mass of the point source arises only
in its angular components).

Alternatively consider (\ref{eq:Phirho}), which may be transcribed to the gravitational context to apply to the infinite cosmic string configuration
within Newtonian gravity, but now to be extended as a solution to General Relativity at least to first-order in the weak field approximation. As is well known
within that context the Newtonian potential provides the first-order correction in $G/c^2$ to the time-time component of the space-time metric.
However given the geometrical meaning of the space-time metric one would require it to be asymptotically flat in all space-like directions, with in particular
its time-time component reaching a constant and finite nonvanishing value, thus requiring an asymptotically vanishing Newtonian potential in all space-like directions.
Of course this is excluded when the source mass distribution extends to spatial infinity in some directions, as is the case with the potential (\ref{eq:Phirho}).
In other words the Newtonian solution (\ref{eq:Phirho}) extends to General Relativity provided only it vanishes, thus producing no gravitational pull
on any time-like geodesic.

Clearly in order to avoid such conclusions it suffices to consider a cosmic dark matter vortex of finite length, thereby doing away at once as well
with translation invariance along the vortex axis, and allowing a nonuniform mass distribution along that axis as well as a varying transverse width
even when still assuming axial symmetry of the gravitational configuration whether in the General Relativity context or the weak field Newtonian
gravity approximation.

This is where the main motivation for the present discussion lies, namely an invitation to consider in earnest the prospects offered by such a hypothesis
towards an understanding of the physical phenomenon responsible for the existence of flat rotation curves for spiral disk galaxies, thus involving
a form of dark matter radically different from any possible type of hitherto undiscovered dark matter particles. The analysis hereafter provides
an initial simple minded such exploration using a rather naive setting, to be pursued and elaborated on much more fully
through more realistic modelling and in a number of complementary directions, both phenomenological and theoretical in character.

The paper is structured as follows. Section~\ref{Sect2} considers some qualitative yet universal characteristics of the Newtonian gravitational potential,
of its gravity field, hence of the associated rotation curve, produced by some arbitrary axially symmetric mass distribution of finite extent,
and invariant under reflection in some point belonging to its symmetry axis.
In section~\ref{Sect3} more specific results are established on basis of a simple and naive model for a vortex's mass distribution.
Yet they allow for some general characteristics of the rotation curve of a vortex to be identified, independent of the details of its mass distribution.
Section~\ref{Sect4} then combines these considerations with an as simple and naive model for a rotating disk galaxy mass distribution, to understand what
the presence of a threading vortex at its centre implies for the overall features of its rotation curve.
Finally section~\ref{Sect5} addresses some concluding remarks and prospects for further work on the potential relevance of dark matter vortices
to the conundrum of dark matter in the Universe.

\section{Universal Qualitative Characteristics of the Gravity Field of an Axially Symmetric Mass Distribution}
\label{Sect2}

Given a static (or stationary) volume mass distribution $\mu(\vec{x}\,)$, the associated Newtonian gravitational potential $\Phi(\vec{x}\,)$ is determined
from the following Poisson equation, with vanishing limiting values at spatial infinity,
\begin{equation}
\vec{\nabla}^2\Phi(\vec{x})=4\pi\,G\,\mu(\vec{x}),\quad
\lim_{|\vec{x}|\rightarrow\infty}\Phi(\vec{x})=0,
\end{equation}
where $G$ is Newton's constant, $G\simeq 4.3\times 10^{-6}\,{\rm kpc}\cdot M^{-1}_\odot\cdot\left({\rm km/s}\right)^2$.

The point mass potential $1/|\vec{x}-\vec{x}'|$ defines effectively the corresponding Green's function (see Appendix A).
Hence the unique solution to this Newton-Poisson equation is provided by the following well-known integral representation,
\begin{equation}
\Phi(\vec{x})=-G\,\int_{(\infty)} d^3\vec{x}\,'\,\frac{1}{|\vec{x}-\vec{x}'|}\,\mu(\vec{x}\,'),
\label{eq:Phi0}
\end{equation}
it being understood that the mass distribution is confined to a volume of finite extent outside of which $\mu(\vec{x})=0$.
In passing, let us point to the linearity in the source $\mu(\vec{x})$ of the Newton-Poisson equation and its solution.

Since the mass distributions to be considered presently are all chosen to be axially symmetric with a common symmetry axis,
let us align that axial symmetry axis with the cartesian coordinate axis $z$, and use cylindrical coordinates $(\rho,\phi,z)$
with $(\rho,\phi)$ being polar coordinates in the transverse plane,
in which case one has $\mu(\vec{x})=\mu(\rho,z)$ without any dependency in the azimuthal angle $\phi$.

Furthermore, in order to better understand the behaviour in the radial transverse distance $\rho$ of the potential $\Phi(\rho,\phi,z)$ and of its derived gravitational field,
$\vec{g}(\rho,\phi,z)=-\vec{\nabla}\Phi(\rho,\phi,z)$, which directly determines the rotation curve for circular trajectories in the transverse $z=0$ plane,
it is advisable to use a representation in which the cylindrical coordinates are separated,
{\it i.e.}, use a representation of the Green's function in terms of a complete set of separated eigenfunctions of the Euclidean Laplacian
adapted to the cylindrical geometry, namely (see Appendix A),
\begin{equation}
\frac{1}{|\vec{x}-\vec{x}'|} = \frac{4}{\pi}\int_0^\infty dk \cos k(z-z')\left\{\frac{1}{2}I_0(k\rho_<) K_0(k\rho_>) +\sum_{m=1}^\infty \cos m(\phi-\phi') I_m(k\rho_<) K_m(k\rho_>) \right\},
\end{equation}
where $\rho_<={\rm min}(\rho,\rho')$, $\rho_>={\rm max}(\rho,\rho')$, and provided that $\rho\ne\rho'$.

Note that the integer $m\in\mathbb{Z}$ in this expression has the
interpretation of a quantised angular momentum measured along the $z$ axis, while the wave number $k\ge 0$ in the $z$ direction provides a direct
harmonic analysis of the mass distribution in its $z$ dependency. It is thus through that wave number in $z$ -- which is a measure of the inverse spatial
extent of the mass distribution and its variations along the $z$ axis -- that the transverse $\rho$ dependency of the potential, hence of the gravitational field
and eventually of the velocity curve,
is directly correlated with the distribution along the $z$ axis of the mass configuration of the system through the contributions in the modified Bessel functions
$I_m(u)$ and $K_m(u)$, $m\ge 0$, in the above representation.

More explicitly, on account of the axial symmetry of the mass distribution, upon integration over the azimuthal angle $\phi'$ the integral representation
for the potential -- which itself is then axially symmetric as well -- reduces to the single zero angular momentum contribution with $m=0$ to the Green's function,
\begin{equation}
\Phi(\rho,z)=-4G\int_0^\infty dk\,\int_0^\infty d\rho'\,\rho'\,\int_{-\infty}^{+\infty}dz'\,\mu(\rho',z')\,\cos k(z-z')\,I_0(k\rho_<)\,K_0(k\rho_>).
\end{equation}
This form for the Newtonian gravitational potential makes it already explicit that beyond the outer radial edge in $\rho$ of the mass distribution
it is the function $K_0(k\rho)$ which governs the radial dependency of the potential, hence of the rotation curve, given a range of values
for the wave number $k$ that would dominate the integral over that parameter.

Since the function $K_0(u)$ diverges as $(-\ln u)$ for small
values of $u>0$, and behaves asymptotically as $e^{-u}\sqrt{\pi/(2u)}$ for large values of $u>0$, clearly the larger is the extent in $z$ of the mass distribution,
the smaller is the range of $k$ values close to $k=0$ which dominate the $k$ integral, and the larger is the range of values in $\rho$ for which
the variation of the gravitational potential through the factor in $K_0(k\rho)$ remains logarithmic in $\rho$ beyond the radial edge in $\rho$ of the mass distribution,
until eventually the exponential decrease of the function $K_0(u)$ sets in for sufficiently large values of $\rho$. In other words there is a direct correlation between
the scale of the extent along the axial symmetry $z$ axis of the mass distribution and the scale of the extent in the range of transverse $\rho$ values
for which the potential remains dominated by a logarithmic dependency in $\rho$, with these two length scales thus being qualitatively essentially of the same order
of magnitude. While such a logarithmic dependency in the potential directly implies a flat rotation curve for circular trajectories in a transverse plane.

Note well that these qualitative conclusions are reached whatever the actual details of the behaviour in $(\rho,z)$ of the mass distribution $\mu(\rho,z)$.
These conclusions directly follow quite generally from the axial symmetry of the mass distribution of finite spatial extent in all directions. They are thus of a universal
character. In particular, qualitatively the length of the flat plateau in the velocity curve is thus a direct order of magnitude ``mirror image'' of the length scale
of the extent along the symmetry $z$ axis of the mass distribution, independently of the details in the mass distribution.
While of course the value of the flat plateau speed itself remains determined by the absolute magnitude of the mass distribution profile in $(\rho,z)$.

In order to pursue with similar qualitative considerations in as simple a manner as possible, let us now assume further that the axially symmetric mass
distribution is symmetric relative to the plane $z=0$, namely that the function $\mu(\rho,z)$ is even in $z$, reflecting a $z$-parity invariance in $z=0$ of
the mass distribution. The $z'$ integration in the above integral representation of the gravitational potential then reduces to,
\begin{equation}
\Phi(\rho,z)=-8G\int_0^\infty dk\,\cos kz\,\int_0^\infty d\rho'\,\rho'\,\int_0^\infty dz'\,\mu(\rho',z')\,\cos kz'\,I_0(k\rho_<)\,K_0(k\rho_>).
\end{equation}
In particular when $\rho$ lies beyond the radial outer edge in $\rho$ of the mass distribution in the transverse plane for a given value of $z$
this expression separates as,
\begin{equation}
\rho>\rho_{\rm outer\ edge}(z):\qquad
\Phi(\rho,z)=-8G\int_0^\infty dk\,\tilde{\mu}(k)\,\cos kz\,K_0(k\rho),
\end{equation}
where the $k$-transformed mass distribution $\tilde{\mu}(k)$ is defined as (note that $\tilde{\mu}(k=0)=M/(4\pi)$ where $M$ is the total mass
of the distribution),
\begin{equation}
\tilde{\mu}(k)=\int_0^\infty d\rho\,\rho\,\int_0^\infty dz\,\mu(\rho,z)\,\cos kz\,I_0(k\rho).
\end{equation}
Both these last two expressions for $\Phi(\rho,z)$ thus show explicitly that in the case of a $z$-parity invariant
axially symmetric mass distribution of finite extent along its symmetry axis,
because of the factor in $\cos kz$, in the direction parallel to that axis the gravitational potential consists of a gravitational well of finite extent in $z$ -- whose range
is determined by the set of $k$ values
for which $\tilde{\mu}(k)$ remains nonnegligible --, centered and symmetric around the plane at $z=0$, and whose minimum sits precisely on that plane
(this longitudinal potential well is deformed when $\mu(\rho,z)$ is not $z$-parity invariant).
As a consequence, quite generally bound trajectories
are expected to coalesce into an accretion disk of finite thickness along $z$, of arbitrary transverse radius along $\rho$, parallel to the plane $z=0$
and centered around it.
This expected qualitative behaviour is yet another universal characteristic of the gravity field of such a mass distribution, and of course directly
reminiscent of the existence of rotating galactic disks in the case of spiral galaxies.

Pursuing further with the $z$-parity invariant axially symmetric mass distribution, let us now turn to its gravitational field expressed as,
\begin{equation}
\vec{g}(\rho,\phi,z)=-\vec{\nabla}\Phi(\rho,z)=-\hat{e}_\rho(\rho,\phi)\,\frac{\partial\Phi(\rho,z)}{\partial \rho} - \hat{e}_z\,\frac{\partial\Phi(\rho,z)}{\partial z}.
\end{equation}
Clearly in view of the above expressions for $\Phi(\rho,z)$ which apply under such conditions, the $z$ component of $\vec{g}(\rho,\phi,z)$ vanishes in the plane $z=0$,
where the gravitational field is thus purely transversely radial. Consequently, on account of nonrelativistic Newtonian dynamics and the centripetal
acceleration $v^2(\rho)/\rho$, the velocity profile $v(\rho)$ of circular trajectories in the plane $z=0$ is such that
\begin{equation}
v^2(\rho)=\rho\frac{\partial\Phi(\rho,z=0)}{\partial\rho}=
-8G\,\rho\frac{d}{d\rho}\int_0^\infty dk\,\int_0^\infty d\rho'\,\rho'\,\int_0^\infty dz'\,\mu(\rho',z')\,\cos kz'\,I_0(k\rho_<)\,K_0(k\rho_>),
\label{eq:v2}
\end{equation}
in which the following identities may be used, $d I_0(k\rho)/d\rho = k\,I_1(k\rho)$ and $K_0(k\rho)/d\rho=-k\,K_1(k\rho)$.
In particular beyond the outer edge of the mass distribution in the $z=0$ plane one has,
\begin{equation}
\rho>\rho_{\rm outer\ edge}(z=0):\qquad
v^2(\rho)=8G\int_0^\infty dk\,\tilde{\mu}(k) \cdot (k\rho) K_1(k\rho).
\end{equation}
Note that $\lim_{u\rightarrow 0^+} u K_1(u)=1$. Hence there exists a genuine possibility indeed for the rotation curve to display an essentially flat section
for a certain range of values in $\rho$ depending on the interval of values for the wave number $k$ around $k=0$ that effectively contribute to the integral.

In order to gain further insight into the properties encoded in these results, let us now turn to specific models of mass distributions,
albeit naive and crude ones sufficient for our present purposes. Note that within the Newtonian weak field approximation to lowest order,
the results for general superpositions of axially symmetric distributions may be obtained through linear superpositions of the corresponding gravitational potentials.

\section{A Crude Model for a Dark Matter Vortex}
\label{Sect3}

First let us aim to identify generic features that the gravitational field specifically of a cosmic dark matter vortex of finite axial extent may possess,
by considering a simple enough, thus naive and crude model for its mass distribution within the context of such an initial exploratory assessment.
For this purpose we shall consider a straight solid cylinder of height $2L$, radius $a$, and uniform mass density $\mu_a$,
with the implicit understanding of an aspect ratio such that $L/a\gg 1$.

In wanting to apply the general results established in the previous section, let us first consider the situation when $\rho>a$, namely for the rotation curve outside
the vortex. By relying on the identity $u\, I_0(u)=d\left(u\,I_1(u)\right)/du$,
the $k$-transformed mass distribution is readily found to be given as (of course with $-L\le z\le L$),
\begin{equation}
\tilde{\mu}(k)=\frac{1}{4\pi}\,\mu_a\cdot\pi a^2\cdot 2L\cdot\frac{\sin kL}{kL}\cdot\frac{I_1(ka)}{ka/2}.
\end{equation}
Note that $\lim_{u\rightarrow 0^+} \sin u/u=1$ and $\lim_{u\rightarrow 0^+} I_1(u)/(u/2)=1$.
Hence indeed $\tilde{\mu}(k=0)=M_a/(4\pi)$ with $M_a=\mu_a\cdot \pi a^2\cdot 2L$ being the total mass of the vortex.

Consequently outside the vortex the rotation curve is expressed through the following integral representation,
\begin{eqnarray}
v^2_a(\rho>a) &=& G\cdot\mu_a\cdot\pi a^2\cdot 2L\cdot\frac{2}{\pi}\int_0^\infty dk\,\frac{\sin kL}{kL}\cdot\frac{I_1(ka)}{(ka/2)}\cdot (k\rho) K_1(k\rho) \nonumber \\
\label{eq:v2_a1}
&=& G\cdot\mu_a\cdot\pi a^2\cdot 2L\cdot\frac{2}{\pi L}\int_0^\infty dx\,\frac{\sin x}{x}\cdot\frac{I_1(\frac{a}{L}x)}{(\frac{a}{L}\cdot\frac{x}{2})}
\cdot\left(\frac{\rho}{L} x\right)K_1(\frac{\rho}{L}x) \\
&=&  G\cdot\mu_a\cdot\pi a^2\cdot 2L\cdot\frac{2}{\pi L}\cdot\frac{2\rho}{a}\int_0^\infty dx\,\frac{\sin x}{x}\cdot I_1(\frac{a}{L}x)\cdot K_1(\frac{\rho}{L}x). \nonumber
\end{eqnarray}
Note how the second of these expressions shows that the scale $L$ of the axial extent of the vortex sets the scale for the relative measurement of all other length
scales in this gravitational system, with the dimensionless variable $x=kL$ playing the role of a modulation and convolution parameter for which the presence
of the factor
$\sin x/x$ implies a leading contribution to the integral from the interval $x\in[0,\pi]$, namely $k\in[0,\pi/L]$ . Incidentally one also has,
\begin{equation}
\lim_{x\rightarrow 0^+}\frac{\sin x}{x}\cdot\frac{I_1(\frac{a}{L}x)}{(\frac{a}{L}\cdot\frac{x}{2})}\cdot(\frac{\rho}{L}x) K_1(\frac{\rho}{L}x) = 1,
\end{equation}
while for large values of $x\rightarrow+\infty$ one has the asymptotic behaviour,
\begin{equation}
\frac{2\rho}{a}\cdot I_1(\frac{a}{L}x)\cdot K_1(\frac{\rho}{L}x)\simeq \frac{L}{a}\sqrt{\frac{\rho}{a}}\cdot\frac{1}{x}\cdot e^{-x(\rho-a)/L}.
\end{equation}
Hence indeed the integral for $v^2_a(\rho>a)$ converges in spite of the exponential increase with $x$ of the function $I_1(ax/L)$ (which is tamed
by the yet stronger exponential decrease in $x$ of $K_1(\rho x/L)$ when $\rho>a$).

Before considering the behaviour in $\rho$ of $v^2_a(\rho>a)$, let us turn to the evaluation of the rotation curve inside the vortex, for $\rho<a$.
The relevant expression is then that of (\ref{eq:v2}), beginning with the integral
\begin{equation}
\int_0^a d\rho'\,\rho'\,I_0(k\rho_<)\,K_0(k\rho_>)=\int_0^\rho d\rho'\,\rho'\,I_0(k\rho')\,K_0(k\rho) + \int_\rho^a d\rho'\,\rho'\,I_0(k\rho)\,K_0(k\rho').
\end{equation}
By relying on the identities $u I_0(u)=d(u I_1(u))/du$ and $u K_0(u) = - d(u K_1(u))/du$,
it readily follows that
\begin{equation}
\int_0^a d\rho'\,\rho'\,I_0(k\rho_<)\,K_0(k\rho_>)=\frac{\rho}{k}\left(I_1(k\rho) K_0(k\rho) + I_0(k\rho) K_1(k\rho)\right)
\,-\,\frac{a}{k} K_1(ka) I_0(k\rho).
\end{equation}
However the velocity profile in (\ref{eq:v2}) requires the evaluation of the $\rho$ derivative of this last expression. Using the identities
\begin{equation}
I'_0(u)=I_1(u),\quad I'_1(u)=I_0(u)-\frac{1}{u}I_1(x),\quad
K'_0(u)=-K_1(u),\quad K'_1(u)=-K_0(u)-\frac{1}{u}K_1(u),
\end{equation}
one finds,
\begin{equation}
\frac{d}{d\rho}\left[\frac{\rho}{k}\left(I_1(k\rho) K_0(k\rho) + I_0(k\rho) K_1(k\rho)\right)\right]=0,\quad
\frac{d}{d\rho}\left[-\frac{a}{k} K_1(ka) I_0(k\rho) \right]=-a K_1(ka) I_1(k\rho).
\end{equation}
Consequently, upon completion of the evaluation of (\ref{eq:v2}), the rotation curve for $\rho<a$ is represented by
\begin{eqnarray}
v^2_a(\rho<a) &=& 8G\cdot \mu_a\cdot \rho L\cdot\int_0^\infty dk \frac{\sin kL}{kL}\cdot a K_1(ka) I_1(k\rho) \nonumber \\
\label{eq:v2_a2}
&=& G\cdot \mu_a\cdot \pi a^2\cdot 2L\cdot\frac{2}{\pi L}\cdot\frac{2\rho }{a}\int_0^\infty dx\,\frac{\sin x}{x}\cdot K_1(\frac{a}{L}x)\cdot I_1(\frac{\rho}{L}x).
\end{eqnarray}
Note how this result for $v^2_a(\rho<a)$ compares with that in (\ref{eq:v2_a1}) for $v^2_a(\rho>a)$, with the interchanges
\begin{equation}
I_1(\frac{a}{L}x) \longleftrightarrow K_1(\frac{a}{L}x),\qquad
K_1(\frac{\rho}{L}x) \longleftrightarrow I_1(\frac{\rho}{L}x),
\end{equation}
mapping one expression into the other (or equivalently under the exchange $a\longleftrightarrow \rho$).
In particular this feature also establishes the continuity of the rotation curve $v_a(\rho)$
at the edge of the vortex at $\rho=a$. Furthermore, the convergence of the integral in (\ref{eq:v2_a2}) when $\rho<a$ may be confirmed along
lines similar to those used previously in establishing the convergence of (\ref{eq:v2_a1}).

Nonetheless because of the wild exponential behaviour for large values of their arguments
in each of the factors in $I_1(u)$ and $K_1(u)$ contributing to these integrals, these representations
of the rotation curve do not easily lend themselves to numerical evaluations. An alternative representation must be found, which no longer relies
on the separation of variables in cylindrical coordinates. The relevant identity reads as follows, with $\rho\ne \rho'$,
\begin{equation}
\int_0^\infty dk\,\frac{\sin kL}{k}\,I_1(k\rho_<)\,K_1(k\rho_>)=\frac{\pi}{2}\int_0^{2\pi} \frac{d\phi}{2\pi} \cos\phi\cdot
\asinh\left(\frac{L}{\sqrt{\rho^2+{\rho'}^2-2\rho\rho'\,\cos\phi}}\right),
\end{equation}
as established in Appendix A, see (\ref{eq:Identity}).

By applying it to the above results for $v^2_a(\rho)$ whether $\rho>a$ or $\rho<a$,
one derives the following dual integral representation of the rotation curve,
now valid for whatever the value of $\rho\ne a$, whether inside or outside the cylindrical vortex,
\begin{equation}
\rho\ne a:\qquad
v^2_a(\rho)=G\cdot \mu_a\cdot \pi a^2\cdot\frac{4\rho}{a}\cdot\int_0^{2\pi}\frac{d\phi}{2\pi}\,\cos\phi\,\asinh\left(\frac{L}{\sqrt{\rho^2+a^2-2a\rho\cos\phi}}\right),
\label{eq:v2_a}
\end{equation}
which proves indeed to possess excellent numerical convergence properties. Once again note the symmetry under the exchange $a\leftrightarrow \rho$
of the r.h.s. of this expression, which is therefore continuous even at $\rho=a$ even though it possesses a cusp-like behaviour at that point
because of the sharp discontinuity in the mass distribution at that value of $\rho$ (in the form of a $\delta(\rho - a)$ singularity in $\partial_\rho\mu(\rho,z=0)$).

Incidentally for the present choice of mass distribution the same result for $v^2_a(\rho)$
may directly be derived from the integral representation (\ref{eq:Phi0}) of the potential
through a straightforward calculation, see Appendix B.

The analytic closed form representation (\ref{eq:v2_a}) of $v_a(\rho)$ allows a detailed study of its behaviour in $\rho$. For small values of $\rho/a$ or $a/\rho$
a series expansion in that quantity may be considered, of which the lowest order nonvanishing contribution leads to the following analytic approximations,
\begin{eqnarray}
\rho<a : \qquad v_a(\rho) &\simeq& \left(2G\cdot \mu_a\cdot \pi a^2\right)^{1/2}\cdot \left(1+\frac{a^2}{L^2}\right)^{-1/4}\cdot\frac{\rho}{a}, \nonumber \\
\rho>a : \qquad v_a(\rho) &\simeq& \left(2G\cdot \mu_a\cdot \pi a^2\right)^{1/2}\cdot \left(1+\frac{\rho^2}{L^2}\right)^{-1/4} \\
&\simeq& \left(2G\cdot \mu_a\cdot \pi a^2\right)^{1/2}\cdot \left(1+\frac{a^2}{L^2}\right)^{-1/4}\cdot\left(\frac{1+\frac{a^2}{L^2}}{1+\frac{\rho^2}{L^2}}\right)^{1/4}.
\nonumber
\end{eqnarray}
However it turns out that these analytic expressions provide most excellent numerical representations of the exact analytic result.
Even for values of $\rho$ that differ from $a$ by as little as a relative value of some $10^{-9}$, these analytic approximations differ from the
actual value of $v_a(\rho)$ by a fraction of a per mille at most. It is only very close indeed to the cusp in $v_a(\rho)$ at $\rho=a$ that these lowest order
approximations may not be relied upon. But of course a smoother radial profile in the mass distribution at its outer edge at $\rho=a$
than the present one would smooth out as well the cusp structure in $v_a(\rho)$, ensuring an even better representation of the exact value by these
analytic approximations, which indeed are continuous at $\rho=a$ with the value
\begin{equation}
v_a(\rho=a) \simeq \left(2G\cdot \mu_a\cdot \pi a^2\right)^{1/2}\cdot \left(1+\frac{a^2}{L^2}\right)^{-1/4}.
\end{equation}
Note well that the quality of these approximations remains independent of the value for $L$.

\begin{figure}
\begin{center}
\includegraphics[width=14cm]{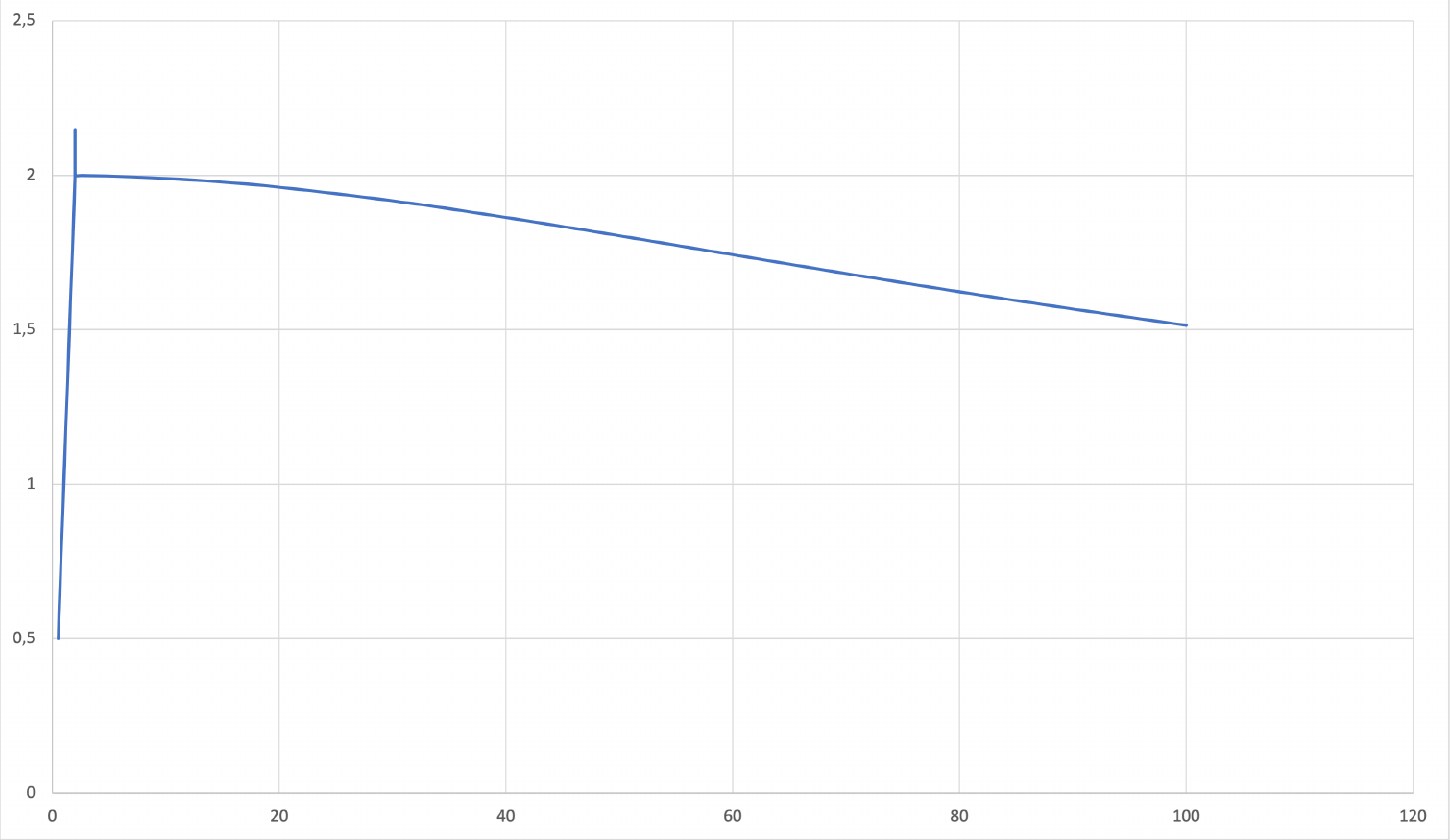}
\end{center}
\caption{The quantity $f(\rho)$ (in units of kpc's) defined in (\ref{eq:frho}), which is directly proportional to the vortex rotation curve $v_a(\rho)$,
as function of $\rho$ (in units of kpc's, horizontal axis) for the choice of parameter values $a=2$~kpc and $L=70$~kpc.
Note the cusp structure due to the sharp boundary of the  cylindrical mass distribution of the vortex at $\rho=a$, as discussed in the text.}
\label{Fig1}
\end{figure}

For illustrative purposes let us introduce the quantity (with a dimension of length) defined by
\begin{equation}
\rho\ne a:\quad
f(\rho)=\sqrt{2a\rho\cdot\int_0^{2\pi}\frac{d\phi}{2\pi}\,\cos\phi\,\asinh\left(\frac{L}{\sqrt{\rho^2+a^2-2a\rho\cos\phi}}\right)}
=\frac{a  v_a(\rho)}{(2G\cdot \mu_a \cdot \pi a^2)^{1/2}},
\label{eq:frho}
\end{equation}
with its lowest order approximations in $\rho/a$ or $a/\rho$ as the case may be,
\begin{equation}
\rho<a:\quad f(\rho) \simeq \rho\, \left(1+\frac{a^2}{L^2}\right)^{-1/4};\qquad\qquad
\rho>a:\quad f(\rho) \simeq a\,\left(1+\frac{\rho^2}{L^2}\right)^{-1/4} .
\label{eq:frho_approx}
\end{equation}
The function $f(\rho)$ (with values in units of kpc's) is displayed in Figure~\ref{Fig1}, for the choice of parameter values $a=2$~kpc and $L=70$~kpc.
This graph corroborates explicitly the above considerations regarding the rotation curve $v_a(\rho)$, inclusive of the cusp structure at $\rho=a$,
the linear variation in $\rho$ for $\rho<a$ up to very close to the cusp, and the very smooth variation in $(1+\rho^2/L^2)^{-1/4}$ for $\rho>a$
starting very close to the cusp (indeed the two analytic approximations of (\ref{eq:frho_approx}) cannot be distinguished numerically from
the exact analytic curve in the graph $f(\rho)$ up to the cusp, which is the reason why they are not displayed with it).

Clearly the linear dependency in $\rho$ of the rotation curve inside the vortex is a direct consequence of the assumed uniformity of its mass distribution.
The details of the profile $v_a(\rho)$ inside the vortex is certainly dependent on the details of its distribution, yet in all cases it should be increasing from a vanishing
value at $\rho=0$ as one moves closer to the radial outer edge of the vortex. In contradistinction, even when the radial outer edge is smoothed out,
one should expect that once sufficiently far outside and away of the vortex's outer edge the rotation curve remains given essentially by the above results,
with a behaviour thus represented to a very good degree of precision by the expression
\begin{equation}
\rho > a:\qquad v_a(\rho)\simeq \sqrt{2G\cdot \lambda_a}\cdot \left(1+\frac{\rho^2}{L^2}\right)^{-1/4},
\end{equation}
where $\lambda_a$ now represents the linear mass density of the vortex along its symmetry axis, with indeed in the case of the present model
$\lambda_a=\mu_a \cdot \pi a^2$. Note how this curve involves only two independent parameters, namely $\lambda_a$ and $L$, with in particular
the length scale for radial decrease in the speed $v_a(\rho)$ set by the symmetric spatial extent $L$ of the vortex along the axial direction of its symmetry axis
perpendicular to the plane of circular orbits.

When $\rho\ll L$, the speed value of the rotation curve in its flat plateau section is thus set by the combination
\begin{equation}
v_{\rm plateau} \simeq \sqrt{2G\cdot \lambda_a},\qquad
\lambda_a \simeq \frac{v^2_{\rm plateau}}{2G}.
\end{equation}
Taking at face value the observation that typically for spiral galaxies $v_{\rm plateau}=v_0\simeq 230$~km/s, this relation translates into
\begin{equation}
\lambda_a\simeq 6.2\times 10^9\ M_\odot \cdot {\rm kpc}^{-1},
\end{equation}
or more generally,
\begin{equation}
150\ {\rm km/s} < v_{\rm plateau} < 300\ {\rm km/s}\quad \longleftrightarrow \quad
2.6\times 10^9\ M_\odot\cdot{\rm kpc}^{-1} < \lambda_a < 10.5\times 10^9\ M_\odot\cdot{\rm kpc}^{-1}.
\end{equation}
These values provide a first estimate (ignoring the contribution of the visible matter of a galaxy to the total rotation curve)
of an order of magnitude for the linear mass density of dark matter vortices, as the dominant drivers of
flat galaxy rotation curves. Note how this conclusion is universal in character, and independent of the (unknown) details of the volume mass distributions
of such vortices.

\section{A Crude Model for a Rotating Disk Galaxy}
\label{Sect4}

Let us now consider the effects on the rotation curve of a dark matter vortex threading the centre of a rotating disk galaxy.
Here as well for such an initial exploratory assessment, we shall consider simple enough, thus naive and crude models
for the different distributions of visible matter composing such a galaxy. While for the vortex itself we use of course still the simple
cylindrical model discussed in section~\ref{Sect3}.

First let us assume that the galaxy harbours a supermassive black hole of mass $M_{\rm BH}$ (when present, a typical value is $M_{\rm BH}\simeq 10^6\ M_\odot$).
At distances sufficiently larger than the scale of its horizon(s), one then has
\begin{equation}
\Phi_{\rm BH}(\rho,z)=-G\frac{M_{\rm BH}}{\sqrt{\rho^2+z^2}},
\end{equation}
thereby providing the following additive contribution to the total rotation curve squared at such distances, in the plane at $z=0$,
\begin{equation}
v^2_{\rm BH}(\rho)=G\frac{M_{\rm BH}}{\rho}.
\end{equation}

The galaxy's bulge is to be modelled as a solid sphere of radius $R_0$ (largely sourrounding the black hole), of total mass $M_{\rm bulge}$,
and of uniform mass density (thus with a sharp boundary as well; typical values are $M_{\rm bulge}\simeq 10^{10}\ M_\odot$).
The corresponding additive contribution to the total rotation curve squared is thus,
\begin{equation}
\rho < R_0:\qquad v^2_{\rm bulge}(\rho)=G\frac{M_{\rm bulge}}{R^3_0}\,\rho^2;\qquad\qquad
\rho > R_0:\qquad v^2_{\rm bulge}(\rho)=G\frac{M_{\rm bulge}}{\rho}.
\end{equation}
Note that for $\rho>R_0$ the total contribution to the rotation curve squared of the black hole and bulge combined is in the form of
\begin{equation}
v^2_0(\rho)=v^2_{\rm BH}(\rho) + v^2_{\rm bulge}(\rho) = G\frac{M_0}{\rho},\qquad
M_0=M_{\rm BH} + M_{\rm bulge}.
\end{equation}
In other words in view of typical values for $M_{\rm BH}$ and $M_{\rm bulge}$, outside the bulge
the possible presence of a supermassive black hole at its centre leads to perfectly negligible effects on the total rotation curve.

The galaxy's disk is to be modelled as a single flat solid circular disk whose symmetry axis is centered and aligned with the $z$ coordinate axis,
of total thickness (or height) $2h$, of radius $b$ with $h\ll b$, of uniform mass density $\mu_b$, and of total mass $M_b$
(typical values being $b \simeq 15$~kpc, $h \simeq 0.1$~kpc, $M_b\simeq 10^{11}\ M_\odot$). In view of the results of section~\ref{Sect3},
the corresponding additive contribution to the total rotation curve squared is,
\begin{equation}
\rho\ne b:\qquad
v^2_b(\rho)=G\cdot\frac{M_b}{2h}\cdot\frac{4\rho}{b}\cdot\int_0^{2\pi}\frac{d\phi}{2\pi}\,\cos\phi\,\asinh\left(\frac{h}{\sqrt{\rho^2+b^2-2b\rho\cos\phi}}\right),
\label{eq:v2_b}
\end{equation}
while beyond the edge of the disk to a very good approximation this expression is numerically equivalent to,
\begin{equation}
\rho>b:\qquad v^2_b(\rho)\simeq G\cdot\frac{M_b}{h}\cdot\frac{1}{\sqrt{1+\frac{\rho^2}{h^2}}}\simeq G\frac{M_b}{\sqrt{\rho^2+h^2}}
\simeq G\frac{M_b}{\rho}\cdot\frac{1}{\sqrt{1+\frac{h^2}{\rho^2}}},
\end{equation}
a result which is independent of the details of the mass distribution inside the disk itself, and thus universal.

Let us also account for a possible spherical stellar and gas halo inside of which the galaxy is embedded, modelled as a solid sphere of radius $R_{\rm sgh}$,
of uniform mass density $\mu_{\rm sgh}$, and of total mass $M_{\rm sgh}$ (typical values being $R_{\rm sgh}\simeq (2-3)\, b$
and $M_{\rm sgh}\simeq 0.1\, M_b$).
Consequently the corresponding additive contribution to the total rotation curve squared is,
\begin{equation}
\rho<R_{\rm sgh}:\qquad v^2_{\rm sgh}(\rho)=G\frac{M_{\rm sgh}}{R^3_{\rm sgh}}\,\rho^2;\qquad\qquad
\rho>R_{\rm sgh}:\qquad v^2_{\rm sgh}(\rho)=G\frac{M_{\rm sgh}}{\rho}.
\end{equation}

And finally there is the contribution of the dark matter vortex modelled as a solid cylinder of uniform mass density $\mu_a$, axial extent $2L$ along the $z$
axis, and radius $a$, which thus contributes with the quantity $v^2_a(\rho)$ in (\ref{eq:v2_a}). For all practical purposes we shall assume that certainly $a<R_0$,
or even $a\ll R_0$, $R_0$ being the spatial extent of the bulge.

Given the linearity of the gravitational potential and field in the mass distribution, the total rotation curve is thus identified from
\begin{equation}
v(\rho)=\sqrt{v^2_0(\rho)+v^2_b(\rho)+v^2_{\rm sgh}(\rho)+v^2_a(\rho)}.
\end{equation}
Relying on the relevant analytic approximations for the values of $v^2_b(\rho)$ and $v^2_a(\rho)$, the following representations apply depending
on the range of values for $\rho$,
\begin{eqnarray}
R_0<\rho<b:\qquad v(\rho) &\simeq& \sqrt{ G\frac{M_0}{\rho} + G\frac{M_b}{\sqrt{b^2+h^2}}\left(\frac{\rho}{b}\right)^2
+G\frac{M_{\rm sgh}}{R^3_{\rm sgh}}\,\rho^2+\frac{2G\cdot \mu_a\cdot \pi a^2}{\sqrt{1+\frac{\rho^2}{L^2}}}  }, \nonumber \\
b<\rho<R_{\rm sgh}:\qquad v(\rho) &\simeq& \sqrt{ G\frac{M_0}{\rho} + G\frac{M_b}{\sqrt{\rho^2+h^2}}
+G\frac{M_{\rm sgh}}{R^3_{\rm sgh}}\,\rho^2+\frac{2G\cdot \mu_a\cdot \pi a^2}{\sqrt{1+\frac{\rho^2}{L^2}}}  }, \\
R_{\rm sgh}<\rho:\qquad v(\rho) &\simeq& \sqrt{ G\frac{M_0}{\rho} + G\frac{M_b}{\sqrt{\rho^2+h^2}}
+G\frac{M_{\rm sgh}}{\rho}+\frac{2G\cdot \mu_a\cdot \pi a^2}{\sqrt{1+\frac{\rho^2}{L^2}}}  }. \nonumber
\end{eqnarray}
Clearly within the range $b<\rho<R_{\rm sgh}$ the contribution to $v^2(\rho)$ of the visible halo proportional to $\rho^2$ helps improve further the
flattening out of the rotation curve in that region of space outside the galaxy and inside its plane, beyond the leading effect provided by the presence
of the vortex. However when combined with the contributions of the bulge and disk whose total masses are larger than $M_{\rm sgh}$ by one to two orders
of magnitude, the contribution due to $M_{\rm sgh}$ remains of second order, and will not be retained for the purpose of the numerical illustration hereafter.

However the numerical simulation to be presented uses the exact analytic expressions for $v^2_b(\rho)$ and $v^2_a(\rho)$ given
in (\ref{eq:v2_b}) and (\ref{eq:v2_a}), and thus computes the rotation curve in the following form,
\begin{equation}
\rho>R_0:\qquad
v(\rho)=\sqrt{v^2_0(\rho)+v^2_b(\rho)+v^2_a(\rho)},\qquad {\rm with}\quad v^2_0(\rho)=\frac{GM_0}{\rho}.
\end{equation}
The contributions to the curve are evaluated with the following (realistic \cite{Sofue2,GRC-MW1}) values of parameters,
\begin{equation}
GM_0=43\,000\ {\rm kpc}\cdot({\rm km/s})^2,\quad
GM_b=70\,000\ {\rm kpc}\cdot({\rm km/s})^2,\quad
G\cdot\mu_a\cdot \pi a^2=26\,000\ ({\rm km/s})^2,
\end{equation}
corresponding to,
\begin{equation}
M_0=10^{10}\ M_\odot,\quad
M_b=1.63\times 10^{10}\ M_\odot,\quad
\mu_a \cdot \pi a^2=6.05\times 10^9\ M_\odot\cdot({\rm kpc})^{-1},
\label{eq:para1}
\end{equation}
as well as,
\begin{equation}
b=10~{\rm kpc},\qquad
h=0.1~{\rm kpc},\qquad
a=1~{\rm kpc},\qquad
L=150~{\rm kpc},
\label{eq:para2}
\end{equation}
while diagrams are drawn for $\rho\ge 4$~kpc assuming that $R_0\simeq 2$~kpc (note that under such circumstances the value for $a$ is of not much import
numerically here, though with $\mu_a\cdot \pi a^2=\lambda_a$ being held fixed). The values for $\rho$ and $v(\rho)$ are then in units of kpc and km/s, respectively.

\begin{figure}
\begin{center}
\includegraphics[width=14cm]{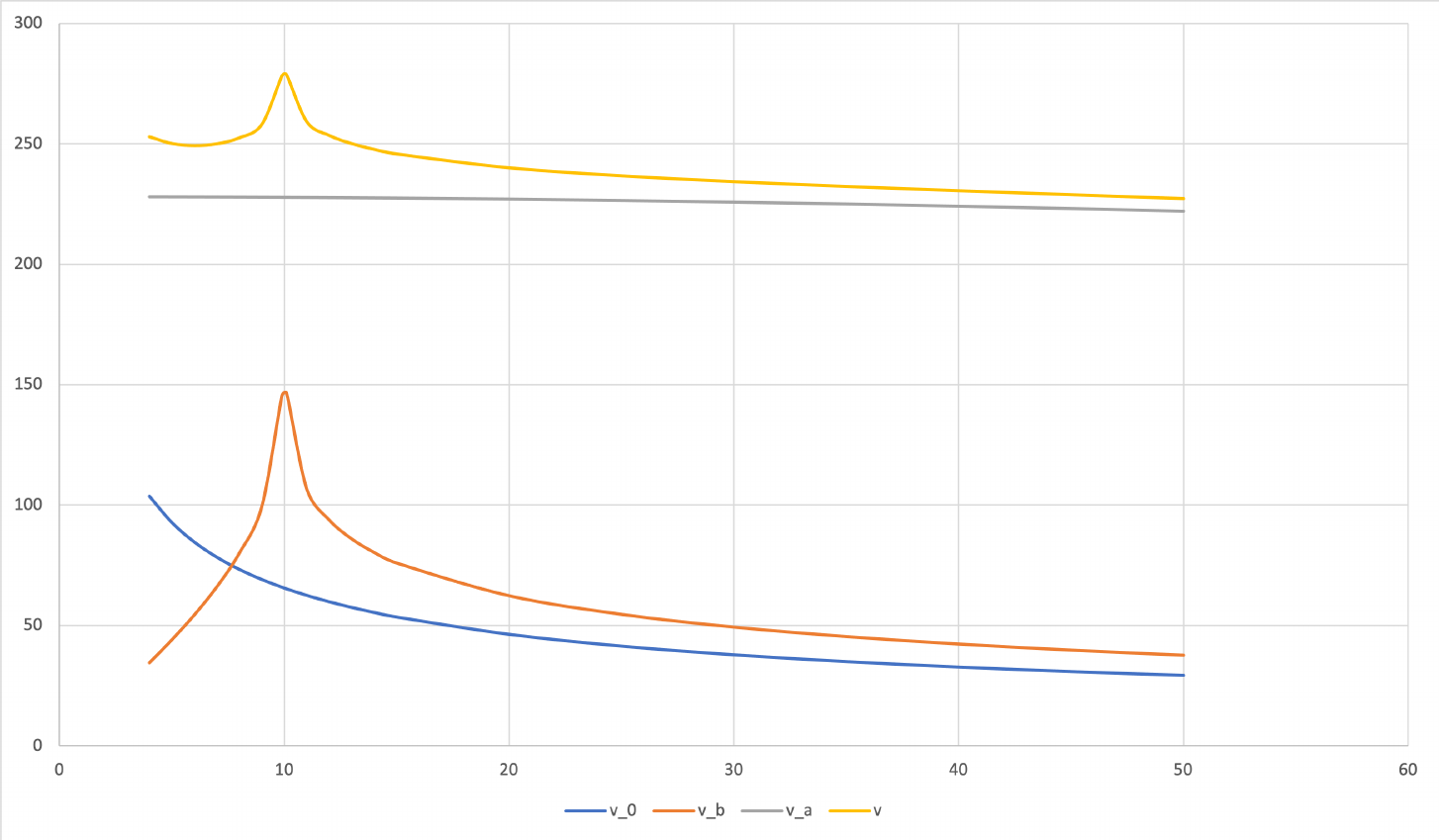}
\includegraphics[width=14cm]{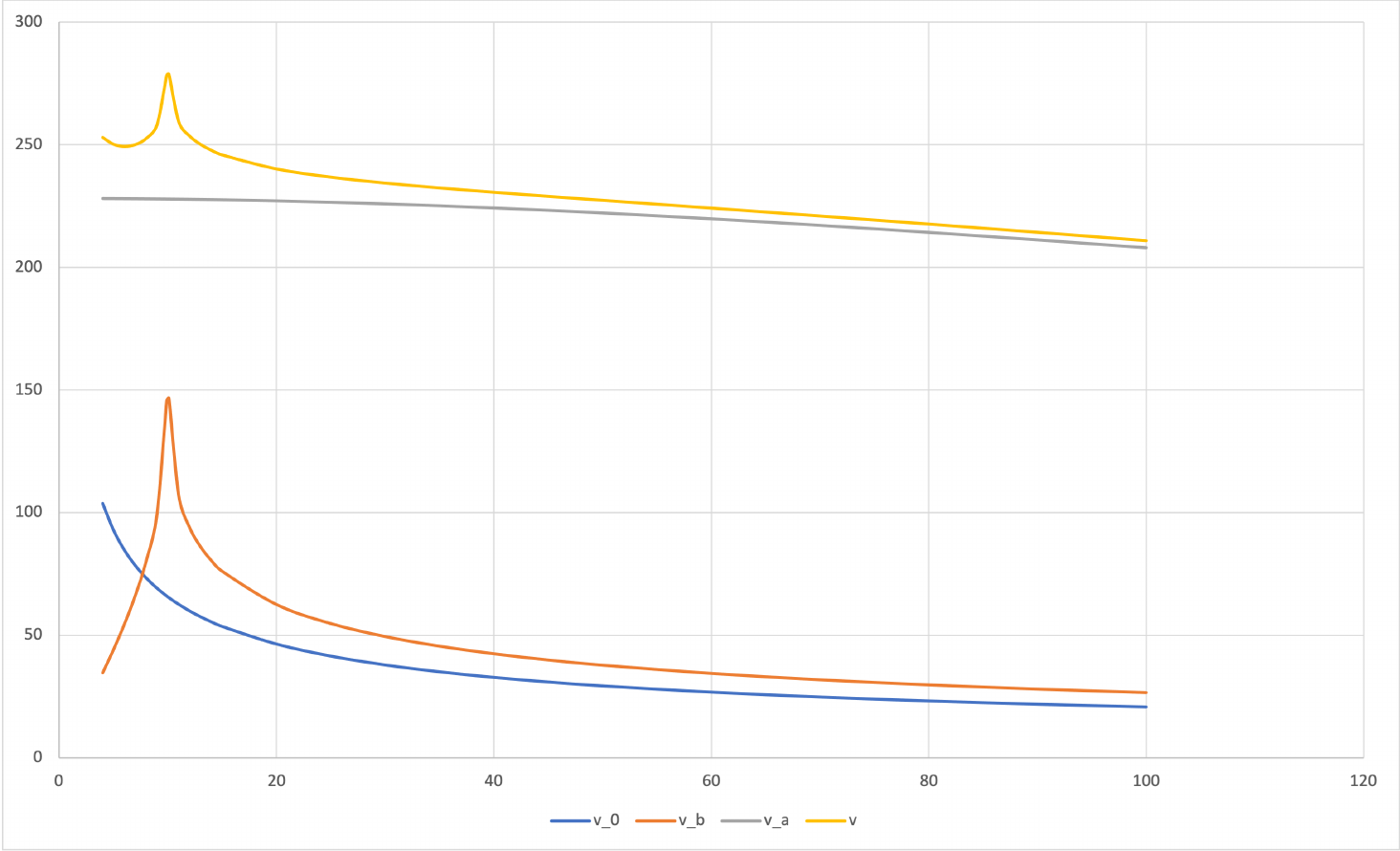}
\end{center}
\caption{From bottom to top in each graph, the rotation curves (in units of km/s) $v_0(\rho)$, $v_b(\rho)$, $v_a(\rho)$ and $v(\rho)$ for
the crude model disk galaxy discussed in section~\ref{Sect4}, as function of $\rho$ (in units kpc's, horizontal axis) with the choice of parameter values
specified in (\ref{eq:para1}) and (\ref{eq:para2}), for $4<\rho<50$ in the upper figure and for $4<\rho<100$ in the lower figure. The cusp
contribution due to $v_b(\rho)$ at $\rho=b=10$~kpc is clearly visible.}
\label{Fig2}
\end{figure}

Figure~\ref{Fig2} displays the resulting values for all four quantities $v_0(\rho)$, $v_b(\rho)$, $v_a(\rho)$ and finally the total rotation curve $v(\rho)$.
The expected cusp at $\rho=b$ due to $v_b(\rho)$ is clearly visible. It would be smoothed out for a smooth radial outer edge of the galaxy disk.
The bulk of the contribution to the total rotation curve is provided by the presence of the vortex through $v_a(\rho)$,
and is thus only very slowly decreasing over tens of kpc's beyond the outer edge of the galaxy at $\rho=b$
on a length scale determined by $L$, and even further beyond that distance (see for instance Figure~\ref{Fig1} for which $L=70$~kpc but $0.5<\rho<100$).

Superposed on the essentially flat rotation curve due to the vortex, further structure essentially confined to the radial extent of the disk galaxy is of course provided
by the details of the visible mass distributions making up the galaxy. The latter may be modelled through realistic models on a case by case basis,
with parameters of such models to be fitted against actually measured rotation curves. In contradistinction, for radial distances outside the volume of the vortex,
its contribution to the total rotation curve involves only two independent parameters, namely its linear mass density $\lambda_a(=\mu_a\cdot \pi a^2)$
and its axial spatial extent scale $L$, with a most simple expression for that contribution,
\begin{equation}
\rho>a:\qquad
v_a(\rho)=\frac{v_{\rm vortex}}{\left(1+\frac{\rho^2}{L^2}\right)^{1/4}},\qquad
v_{\rm vortex}=\sqrt{2G\cdot \lambda_a}.
\end{equation}
Thus $\lambda_a$ sets the scale of the speed value $v_{\rm vortex}$ that produces and dominates the essentially flat plateau in the rotation curve,
while $L$ determines the length scale of the radial extent of that plateau beyond the radial boundary of the galaxy. Only these two parameters are required
for a fit of the vortex's contribution to the total rotation curve. And they readily account for the existence of an
essentially flat rotation curve far beyond the edges of the galaxy, indeed in a convincing and realistic enough way.

\section{Conclusions}
\label{Sect5}

Unsurprisingly, the preliminary simple minded analysis pursued in this paper has established that a flat component to spiral galaxy rotation curves
extending over tens of kpc's beyond the visible outer edge of the disk's galaxy may simply be the manifestation of the presence of an axially symmetric
mass distribution threading the galaxy's centre, perpendicular to its disk, and having a comparable extent along that direction on both sides of the galaxy,
namely a vortex of finite length and comparatively small transverse width.

Some universal characteristics have been identified, independent of the detailed structures of the mass distributions involved. In the direction perpendicular
to the disk the gravitational potential of such a vortex has the profile of a potential well with its minimum inside the disk.
This property would readily lead to the coalescence of
close to circular trajectories into the galaxy's accretion disk. In the radial transverse direction the gravitation potential has a logarithmic behaviour,
readily producing an essentially flat rotation curve. The radial extent of the flat component contribution to the total rotation curve is directly the
``mirror image'' of the extent either side of the disk of the perpendicular vortex itself, say of length $L$. While the vortex's contribution to the total rotation curve involves
only two independent parameters, namely $L$, and additionally the overall scale for the associated speed, $\bar{v}_{\rm vortex}$, in the most simple form of
\begin{equation}
v^2_{\rm vortex}(\rho) \simeq \frac{\bar{v}^2_{\rm vortex}}{\sqrt{1+\frac{\rho^2}{L^2}}},
\end{equation}
or equivalently the characteristic linear mass density of the vortex, $\lambda_{\rm vortex}$,
\begin{equation}
\bar{v}^2_{\rm vortex}\simeq 2G\cdot \lambda_{\rm vortex}.
\end{equation}
A first order of magnitude estimate, based on $\bar{v}_{\rm vortex}\simeq 230$ km/s,
leads to the value $\lambda_{\rm vortex}\simeq 6.2\times 10^9\,M_\odot\cdot{\rm kpc}^{-1}$.
Note that with $2L=300$ kpc, this amounts to a total vortex mass of $M_a=1.86\times 10^{12}\,M_\odot$, a value which compares favourably with the
estimated total mass of the dark matter halo, say, of the Milky Way of some $6\times 10^{11} - 3\times 10^{12}\,M_\odot$ \cite{DMHalo}.

The nature of the vortex's matter composition is unknown and nonvisible, and is thus of a dark matter character. However in contradistinction to standard explanations
for flat rotation curves involving a halo of dark matter particles in which the galaxy is embedded and with a spherically symmetric mass distribution
to be tailored to reproduce the observed flat component to the rotation curve on a case by case basis, such a dark matter vortex -- that makes the galaxy
a component of a rotating top of intergalactic proportions -- provides a totally different alternative towards an understanding of flat rotation curves,
which does not rely on the existence of dark matter particles.

The present work is an invitation to consider in earnest the potential offered by this alternative approach to the dark matter mystery of the Universe.
It may be developed in a number of independent, complementary, and necessary directions.

Of course a {\sl bona fide} analysis of the measured rotation curves of a large sample of spiral galaxies to be fitted based on their known visible matter distributions
in the presence of such a dark matter vortex distribution ought to be performed. As a matter of fact, shortly after this work was first written up in January 2022,
in April 2022 Ref.\cite{Llanes} presented a discussion along somewhat similar considerations precisely based on actual data of galaxy rotation curves.

Incidentally, because of the gravitational potential well aligned with the vortex axis, the issue of the so-called missing satellite 
problem \cite{Satellite1,Satellite2,Satellite3,Satellite4} would deserve to be addressed as well in such a context. Likewise another issue
is that of lensing induced by the vortex, whether in the plane transverse to it, as well as for a source aligned with its symmetry axis.

Such axially symmetric solutions to Newton's gravity may be extended to axially symmetric solutions, including rotating ones,
to the equations of General relativity, thereby also allowing for an analysis of the corresponding relativistic corrections as well as
the corrections implied by rotating sources through gravitoelectromagnetic effects. The time-like geodesics and bound orbits for such space-time geometries should
then also be understood more thoroughly. In fact, motivated by the considerations developed in the present work,
such a study has now been completed at least in the weak-field limit, specifically in the case of the gravimagnetic dipole \cite{Clement}.
This exact solution to Einstein's equations of General Relativity consists of two rotating black holes of identical masses but opposite
Newman-Unti-Tamburino (NUT) charges -- the gravitational analogue of Dirac monopole point charges -- connected
by a Misner string -- the gravitational analogue of a Dirac string. As a linearly distributed singularity in the local space-time curvature yet of vanishing tension,
thus truly dark indeed since of purely gravitational origin and character, the Misner string sourced by the NUT charges
then plays a role akin to that of a dark matter vortex in the nature of the present paper's proposal.
The presence of the NUT charges may then lead to a larger flat plateau in the rotation velocity curve, thus possibly improving on
the purely Newtonian context of the present work's analysis and its conclusions following from the presence of a dark matter vortex \cite{Govaerts}.

The extension of such an alternative to all other instances for which the existence of dark matter has been shown to be critical,
from structure formation to cosmology, must be endeavoured as well.

Finally the possible physical nature of such dark matter vortices must be addressed, beginning with the known physics
of the Standard Model of the electroweak and strong interactions and in particular of its Higgs sector, and then of course beyond that framework.

Of note in that respect are recent results such as those that point possibly to cosmic string effects in some stochastic gravitational
wave signals \cite{CStrings1,CStrings2}.
Or also the growing evidence for the existence of inter-galactic spinning filaments extending outside the core of galaxies, and spanning large webs of cosmic
proportions \cite{Filament1,Filament2,Filament3,Filament4}.

\section*{Appendix A}
\label{AppendixA}

The (Coulomb or Newton) scalar potential $1/|\vec{x}-\vec{x}'|$ is of course the Laplacian Green's function
in 3-d Euclidean space with vanishing limit values at spatial infinity, namely
\begin{equation}
\vec{\nabla}^2_{\vec{x}}\,\frac{1}{|\vec{x}-\vec{x}'|}=-4\pi\,\delta^{(3)}(\vec{x}-\vec{x}'),\qquad
\lim_{|\vec{x}|\rightarrow\infty}\frac{1}{|\vec{x}-\vec{x}'|}=0.
\end{equation}
Given the axial symmetry of matter distributions considered in the present work, it proves useful to consider
the separation of this Green's function in cylindrical coordinates $(\rho,\phi,z)$ in terms of the following
complete basis representation, valid for $\rho\ne\rho'$ \cite{Jackson},
\begin{eqnarray}
\label{eq:Green1}
\!\! \frac{1}{|\vec{x}-\vec{x}'|} &=& \frac{2}{\pi}\sum_{m=-\infty}^\infty\int_0^\infty dk\,e^{im(\phi-\phi')} \cos k(z-z')\,I_m(k\rho_<) K_m(k \rho_>)  \\
&=&\frac{4}{\pi}\int_0^\infty dk \cos k(z-z')\left\{\frac{1}{2}I_0(k\rho_<) K_0(k\rho_>) +\sum_{m=1}^\infty \cos m(\phi-\phi') I_m(k\rho_<) K_m(k\rho_>) \right\},
\nonumber 
\end{eqnarray}
where $I_m(k\rho)$ and $K_m(k\rho)$ are the modified Bessel functions of the first and second kind
of order $m\in\mathbb{Z}$, respectively, while $\rho_<={\rm min}(\rho,\rho')$ and $\rho_>={\rm max}(\rho,\rho')$.
Obviously $(\rho,\phi,z)$ (resp., $(\rho',\phi',z')$) are the cylindrical coordinates of the position vector $\vec{x}$ (resp., $\vec{x}'$).

Note that since $I_0(0)=1$ and $I_{m\ne 0}(0)=0$ one has the following simple identity, obtained by taking $\vec{x}'=\vec{0}$ in the above decomposition,
\begin{equation}
\frac{1}{\sqrt{\rho^2+z^2}}=\frac{2}{\pi}\int_0^k dk \cos kz\, K_0(k\rho),
\end{equation}
while $K_0(x)$ diverges logarithmically at $x=0$.

Since $|\vec{x}-\vec{x}'|=\sqrt{\rho^2+{\rho'}^2-2\rho\rho'\cos(\phi-\phi')+(z-z')^2}$, the representation (\ref{eq:Green1}) implies
the following simple identity as well, valid when $\rho\ne\rho'$,
\begin{equation}
\int_0^{2\pi} d\phi\,\frac{\cos\phi}{\sqrt{\rho^2+{\rho'}^2-2\rho\rho'\cos(\phi-\phi')+(z-z')^2}}=4\int_0^\infty dk \cos k(z-z')\,I_1(k\rho_<)\,K_1(k\rho_>),
\end{equation}
while with $z=0$ (and denoting $z'$ then as $z$),
\begin{equation}
\int_0^{2\pi} d\phi\,\frac{\cos\phi}{\sqrt{\rho^2+{\rho'}^2-2\rho\rho'\cos\phi+z^2}}=4\int_0^\infty dk \cos kz\,I_1(k\rho_<)\,K_1(k\rho_>).
\end{equation}

In particular consider the $z$-integral of the latter relation over the finite interval $[0,L]$ with a uniform distribution, thus with
$\int_0^L dz\,\cos kz=\sin kL/k$ in the r.h.s., so that
\begin{equation}
\int_0^\infty dk\,\frac{\sin kL}{k}\,I_1(k\rho_<)\,K_1(k\rho_>)=\frac{1}{4}\int_0^L dz\,\int_0^{2\pi}d\phi\,\frac{\cos\phi}{\sqrt{\rho^2+{\rho'}^2-2\rho\rho'\cos\phi+z^2}}.
\end{equation}
Finally the $z$-integral in the r.h.s. of this latter relation is straightforward enough to complete, leading to the following identity, valid for $\rho\ne\rho'$,
\begin{equation}
\int_0^\infty dk\,\frac{\sin kL}{k}\,I_1(k\rho_<)\,K_1(k\rho_>)=\frac{\pi}{2}\int_0^{2\pi} \frac{d\phi}{2\pi} \cos\phi \cdot
\asinh\left(\frac{L}{\sqrt{\rho^2+{\rho'}^2-2\rho\rho'\,\cos\phi}}\right).
\label{eq:Identity}
\end{equation}
Note the following alternative representation for the function $\asinh x$, valid for $x\in\mathbb{R}$,
\begin{equation}
\asinh x=\ln\left(x+\sqrt{x^2+1}\right).
\end{equation}

\section*{Appendix B}
\label{Appendix B}

Consider the Poisson equation for the Newtonian gravitational potential $\Phi(\vec{x}\,)$ sourced by a volume mass distribution $\mu(\vec{x})$
and required to vanish at spatial infinity,
\begin{equation}
\vec{\nabla}^2_{\vec{x}}\Phi(\vec{x}\,)=4\pi\,G\,\mu(\vec{x}\,),\qquad
\lim_{|\vec{x}|\rightarrow\infty}\Phi(\vec{x}\,)=0.
\end{equation}
Its solution is thus represented by the integral expression
\begin{equation}
\Phi(\vec{x}\,)=-G\int_{(\infty)}d^3\vec{x}\,'\,\frac{1}{|\vec{x}-\vec{x}'|}\,\mu(\vec{x}\,').
\end{equation}
Using the fact that the mass distribution is assumed to be confined to a volume of finite extent and thus to vanish at spatial infinity, and the property that
$\vec{\nabla}_{\vec{x}}|\vec{x}-\vec{x}'|^{-1}=-\vec{\nabla}_{\vec{x}'}|\vec{x}-\vec{x}'|^{-1}$, the associated Newtonian gravitational field is readily expressed as,
\begin{equation}
\vec{g}(\vec{x}\,)=-\vec{\nabla}_{\vec{x}}\Phi(\vec{x}\,)=G\int_{(\infty)}d^3\vec{x}\,'\,\frac{1}{|\vec{x}-\vec{x}'|}\,\vec{\nabla}_{\vec{x}'}\mu(\vec{x}\,').
\end{equation}

Let us restrict now to an axially symmetric mass distribution, namely $\mu(\rho,z)$ in cylindrical coordinates $(\rho,\phi,z)$, which is also
$z$-parity invariant in $z=0$ ({\it i.e.}, even in $z$, or symmetric relative to the plane $z=0$).
In that case the gravitational field is axially symmetric,
has no azimuthal component, namely $g_\phi(\rho,z)=0$, and is purely radial in $\rho$ in the transverse plane at $z=0$ with as radial component
\begin{equation}
g_\rho(\rho,z=0)=G\int_0^\infty d\rho'\,\rho'\,\int_0^{2\pi} d\phi'\,\int_{-\infty}^{+\infty}dz'\,\frac{\cos\phi'}{\sqrt{\rho^2+{\rho'}^2-2\rho\rho'\cos\phi' + {z'}^2}}\,
\frac{\partial}{\partial\rho'}\mu(\rho',z').
\end{equation}
Consequently in the limit of nonrelativistic Newtonian dynamics the rotation curve for circular trajectories in the plane $z=0$ is given by
\begin{equation}
v^2(\rho)=-\rho\,g_\rho(\rho,z=0).
\end{equation}

In the case that the mass distribution factorises as
\begin{equation}
\mu(\rho,z)=\mu_\rho(\rho)\,\mu_z(z),
\end{equation}
one has
\begin{equation}
v^2(\rho)=-G\rho\,\int_0^{2\pi}d\phi\,\cos\phi\,\int_{-\infty}^{+\infty}dz\,\mu_z(z)\,\int_0^\infty d\rho'\,
\frac{\rho'}{\sqrt{\rho^2+{\rho'}^2-2\rho\rho'\cos\phi+z^2}}\frac{d\mu_\rho(\rho')}{d\rho'}.
\end{equation}

For a cylindrical mass distribution of uniform volume mass density $\mu_0$, radius $a$ and height $2L$, such that,
\begin{equation}
\mu(\rho,z)=\mu_0\,\theta(a-\rho)\left(\theta(z+L)-\theta(z-L)\right),\quad
\frac{\partial}{\partial\rho}\mu(\rho,z)=-\mu_0\,\delta(a-\rho)\left(\theta(z+L)-\theta(z-L)\right),
\end{equation}
where $\theta(x)$ is the usual Heaviside step distribution with $\theta(x>0)=+1$ and $d\theta(x)/dx=\delta(x)$, and $\delta(x)$ being Dirac's $\delta$ distribution,
both integrals over $\rho'$ and $z$ are straightforward enough, leading to the final integral representation for the rotation curve in that case which is valid
for all values of $\rho$ such that $\rho\ne a$,
\begin{equation}
\rho\ne a:\qquad
v^2(\rho)=G\,\mu_0\,2a\rho\,\int_0^{2\pi}d\phi\,\cos\phi \cdot \asinh\left(\frac{L}{\sqrt{\rho^2+a^2-2a\rho\cos\phi}}\right).
\end{equation}
Note how the r.h.s.~of this expression is symmetric under the interchange of $\rho$ and $a$, hence continuous at $\rho=a$.

\clearpage

\end{document}